\documentclass[iop,apj]{emulateapj} 
\usepackage[utf8]{inputenc}
\usepackage[T1]{fontenc}
\usepackage[english]{babel}
\usepackage{amsmath}
\usepackage[upint]{stix}
\usepackage{helvet}
\usepackage{graphicx}
\usepackage[]{lineno}

\newcommand{\td}[1]{\, \mbox{d} #1 \,}
\newcommand{\intl}{\int\limits}

\newcommand{\HF}[1]{\; H\left[ #1 \right]}  
\newcommand{\DF}[1]{\; \delta\left( #1 \right)}  
\newcommand{\fermi}{\textit{Fermi}-LAT\,}
\newcommand{\g}{\ensuremath{\gamma}}
\newcommand{\cta}{\object{CTA\,102}\,} 
\newcommand{\zred}{z_{\rm red}}

\newcommand{\logb}[1]{\ln{\left( #1 \right)}}
\newcommand{\p}{^{\prime}}

\begin{document}

\shorttitle{Cloud ablation in \cta}
\shortauthors{M. Zacharias, et al.}

\title{Cloud ablation by a relativistic jet and the extended flare in \cta in 2016 and 2017}

\author{M. Zacharias$^1$, M. B\"ottcher$^1$, F. Jankowsky$^2$, J.-P. Lenain$^3$, S.J. Wagner$^2$, A. Wierzcholska$^4$}
\email{mzacharias.phys@gmail.com}
\affiliation{$^1$Centre for Space Science, North-West University, Potchefstroom, 2520, South Africa \\
$^2$Landessternwarte, Universit\"at Heidelberg, K\"onigstuhl, D-69117 Heidelberg, Germany \\
$^3$Sorbonne Universit\'es, UPMC Universit\'e Paris 06, Universit\'e Paris Diderot, Sorbonne Paris Cit\'e, CNRS, Laboratoire de Physique Nucl\'eaire et de Hautes Energies (LPNHE), 4 place Jussieu, F-75252, Paris Cedex 5, France \\
$^4$Institute of Nuclear Physics, Polish Academy of Sciences, PL-31342 Krakow, Poland
}

\begin{abstract}
In late 2016 and early 2017 the flat spectrum radio quasar \cta exhibited a very strong and long-lasting outburst. The event can be described by a roughly 2 months long increase of the baseline flux in the monitored energy bands (optical to $\gamma$ rays) by a factor 8, and a subsequent decrease over another 2 months back to pre-flare levels. The long-term trend was superseded by short but very strong flares, resulting in a peak flux that was a factor 50 above pre-flare levels in the \g-ray domain and almost a factor 100 above pre-flare levels in the optical domain. 
In this paper we explain the long-term evolution of the outburst by the ablation of a gas cloud penetrating the relativistic jet. The slice-by-slice ablation results in a gradual increase of the particle injection until the center of the cloud is reached, after which the injected number of particles decreases again. With reasonable cloud parameters we obtain excellent fits of the long-term trend. 
\end{abstract}


\maketitle
%
%
\section{Introduction}
Blazars, the relativistically beamed, radio-loud version of active galactic nuclei \citep{br74}, are historically categorized in two classes depending on the width of their optical emission lines: BL Lacertae objects with line equivalent width ${\rm EW}<5\,$\AA, and flat spectrum radio quasars (FSRQs) with ${\rm EW}>5\,$\AA. The latter case indicates the presence of a strong broad-line region (BLR) surrounding the central supermassive black hole on scales of $\sim 0.1\,$pc. The origin of the double-humped spectral energy distribution (SED) is regarded by most authors to be synchrotron and inverse-Compton (IC) emission of particles within the relativistic jet, with electrons and positrons being responsible for the emission, and protons serving as a cold background. Especially in FSRQs, seed photon fields for the IC process are abundant. Apart from the emission region's internal synchrotron emission (resulting in synchrotron-self Compton, SSC, flux), also the external fields from the accretion disk, the BLR or the dusty torus are potential targets depending on the distance of the emission region from the black hole.

Blazars are strongly variable in all energy bands. The large variety in flaring events has led to a similarly large number of models. A particularly interesting case is the interaction of the jet with an obstacle, such as a star \citep{bk79,k94,pmlh14,b15}, its wind \citep{abr09,dcea17} or a gas cloud \citep{abr10,bpb12}. Most of these models have in common that the obstacle is already fully inside the jet before the start of the interaction. However, given the strong pressure of the relativistically moving matter of the jet, interactions will start as soon as the obstacle hits the jet, since the jet will look like a strong shock. Simulations of shock/cloud interactions have shown that a cloud will be quickly ripped apart \citep{kkc94,pfb02}. Recent simulations of a jet/cloud \citep{bpb12} or jet/star \citep{pbb17} interaction, where the penetration process is included, reveal that the obstacle is (partially) ablated, and a significant amount of matter is mixed into the jet flow.

This is easy to see for a gas cloud, given that it is mainly confined by its own, rather weak gravity. The ram pressure of the jet will immediately start to ablate the outer layers of the cloud while it starts to penetrate the jet. The mass loss of the cloud will weaken its structural integrity even before it has fully penetrated the jet. As we will discuss below, the cloud will be ablated and carried along by the jet. Depending on the cloud parameters, such as size and velocity, this might lead to pronounced and prolonged jet activity, when the additional material in the jet reaches an internal shock located downstream of the cloud penetration site. We apply this model to a recent flare in \object{CTA\,102}, where fluxes varied significantly over several months.

\cta is an FSRQ at a redshift $\zred = 1.037$, roughly half-way across the observable Universe. The accretion disk luminosity is $L_{\rm disk}\p = 3.8\times 10^{46}\,$erg/s \citep{zcst14}. The mass of the central black hole is estimated at $M_{\rm bh}\sim 8.5 \times 10^8\, M_{\odot}$ \citep{zcst14} giving an Eddington luminosity of $L_{\rm Edd}\p\sim 1.1\times 10^{47}\,$erg/s. The BLR properties have been derived by \cite{pft05} using UV spectroscopy observations with the {\it Hubble Space Telescope}, resulting in a luminosity of $L_{\rm BLR}\p = 4.14\times 10^{45}\,$erg/s, and a radius of $R_{\rm BLR}\p = 6.7\times 10^{17}\,$cm (all quantities given in the AGN frame).

Long-term observations in radio bands since 1980 \citep{fea11} revealed a rather dormant source until $\sim$1997, after which it showed a few radio outbursts with a particularly strong one in 2006. \cite{fea11} favor a shock-shock interaction scenario to explain the observed evolution of the latter event. Similarly, in the high energy (HE, $E>100\,$MeV) $\gamma$-ray band, scanned continuously by the {\it Fermi} satellite since mid-2008, \cta showed low fluxes in the first almost four years of \fermi operation with an average flux above $1\,$GeV of $(5.0\pm 0.2)\times 10^{-9}\,$ph/cm$^2$/s and photon index of $\Gamma=2.34\pm 0.03$ \citep{aFea15}. In the second half of 2012 \cta exhibited a strong $\gamma$-ray outburst with a peak flux above $100\,$MeV of $\sim 8\times 10^{-6}\,$ph/cm$^2$/s. This outburst along with correlated optical variability led \cite{lea16} to propose the helical motion and the accompanied variation of the Doppler factor of a plasma blob \citep{sea93} as the main driver of the flare. Since 2012, \cta remained active without long returns to pre-flare levels in both the $\gamma$-ray and optical bands. However, all these outbursts have been rather short lived on the order of a few days, with fast rises to the maximum and subsequent quick decays. 

This behavior changed in late 2016, when \cta entered into a prolonged activity phase, which saw both the $\gamma$-ray and optical fluxes, as well as the X-ray flux rising continuously for about 2 months. The peak fluxes were obtained at the end of December 2016, which were in all cases significantly higher than any previously observed fluxes. The optical fluxes exhibited clear intra-night variability \citep{bea17}. Subsequently, the $\gamma$-ray flux decreased over the course of about 2 months to October 2016 levels. Unfortunately, this decrease of flux could not be observed in optical or X-ray observations due to sun constraints. 

In this paper we present the multiwavelength data of this roughly 4 months long outburst and explain it by the ablation of a gas cloud by the relativistic jet. The initial density increase in ablated material causes the rise of the lightcurve, while the ablation of the second half of the cloud exhibits a decrease in ablated material resulting in the subsequent drop of the lightcurve. 
Our focus is on the explanation of the long-term trend and we do not deal with the fast variability on top of the longer trend.
The paper is organized as follows.
First we present the data analysis in section \ref{sec:ana}. Section \ref{sec:theory} describes the theoretical model of cloud ablation, followed by a summary of the used code and the modeling in section \ref{sec:mod}. We discuss and conclude in section \ref{sec:dis}.

In the following sections, primed quantities are in the AGN frame, quantities marked with the superscript ``obs'' are in the observer's frame, and unmarked quantities are in the comoving jet frame. We use a standard, flat cosmology with $H_0 = 69.6\,$km/s/Mpc, and $\Omega_M = 0.27$, which gives a luminosity distance $d_L = 2.19\times 10^{28}\,$cm.

%
%
\section{Data analysis} \label{sec:ana}
The flare in \cta was extensively observed by a large number of observatories. Here we analyze and report the detailed observations of \fermi in the \g-ray band, Swift-XRT in the X-ray band, as well as Swift-UVOT and ATOM in the optical band.
\begin{figure}[t]
\centering 
\includegraphics[width=0.48\textwidth]{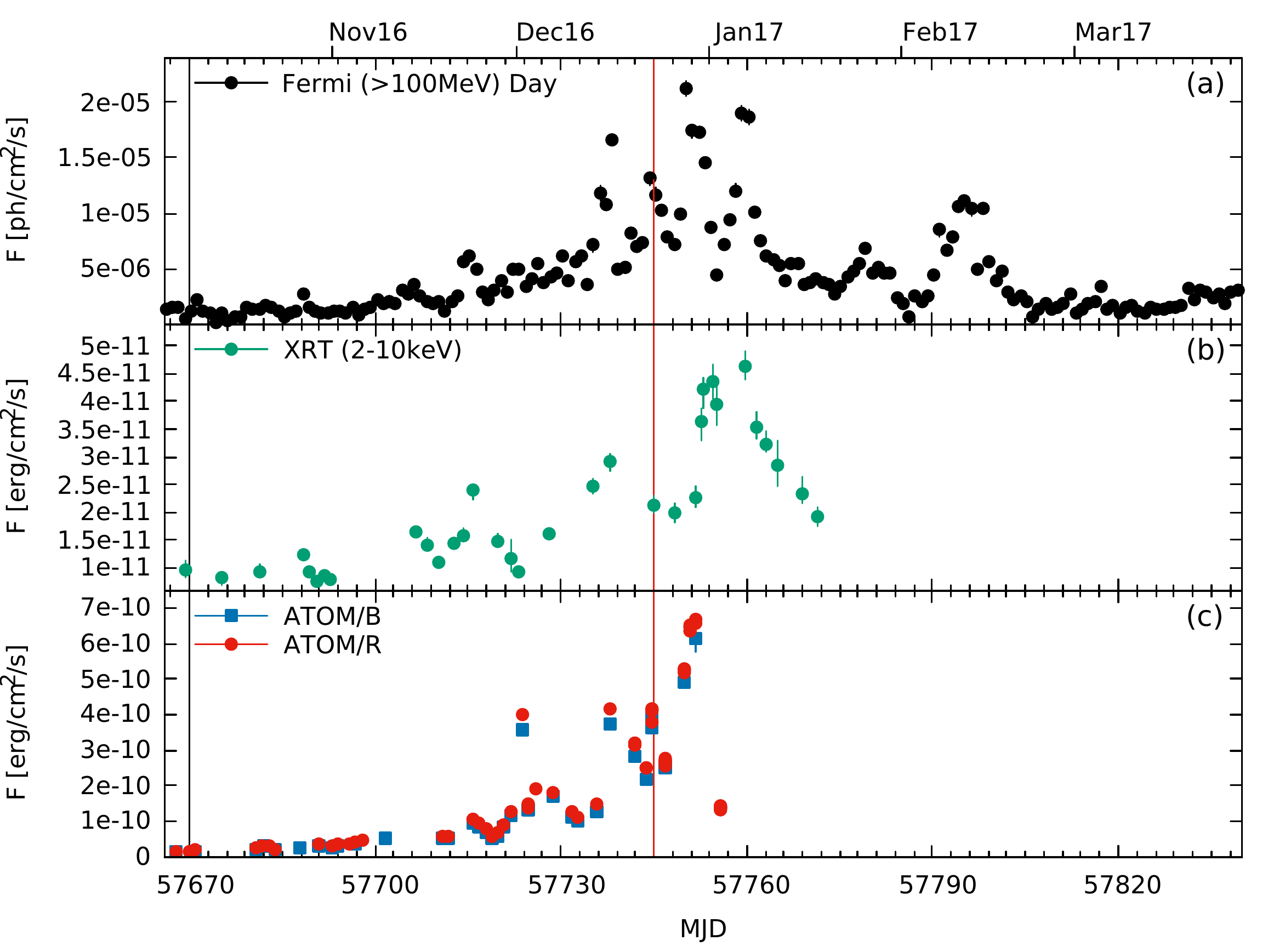}
\caption{Lightcurves of {\bf (a)} \fermi data, {\bf (b)} {\it Swift}-XRT data, and {\bf (c)} optical data from ATOM as labeled. The vertical thin black and red lines mark the dates, where the spectra have been extracted.}
\label{fig:lc}
\end{figure} 
%

\subsection{Fermi-LAT data analysis} \label{sec:fermi}
The LAT instrument \citep{aFea09} onboard the \textit{Fermi} satellite surveys the high energy \g-ray sky every 3 hours, with energies between 20\,MeV and above 300\,GeV, thus making it an ideal instrument to monitor the activity of \cta. This AGN has been reported in all the available \fermi catalogs, and is identified as \object{3FGL~J2232.5+1143} in the third \fermi source catalog \citep{aFea15}.

The \fermi data are analyzed using the public ScienceTools \texttt{v10r0p5}\footnote{See \href{http://fermi.gsfc.nasa.gov/ssc/data/analysis/documentation}{http://fermi.gsfc.nasa.gov/ssc/data/analysis/documentation}.}. Events in a circular region of interest of 10\degr\ in radius are extracted, centered on the nominal position of \object{3FGL~J2232.5+1143}. To probe the active state reported here, only data between August 8, 2015 (MJD 57235) and May 1, 2017 (MJD 57874) are considered, in the 100\,MeV--500\,GeV energy range. The \texttt{P8R2\_SOURCE\_V6} instrument response functions (event class 128 and event type 3) were used, together with a zenith angle cut of 90\degr\ to avoid contamination by the \g-ray bright Earth limb emission. The model of the region of interest was built based on the 3FGL catalog \citep{aFea15}. The Galactic diffuse emission has been modeled using the file \texttt{gll\_iem\_v06.fits} \citep{aFea16} and the isotropic background using \texttt{iso\_P8R2\_SOURCE\_V6\_v06.txt}. In the following, the source spectrum will be investigated both with a power-law shape

\begin{equation}
  \frac{dN}{dE} = N_0 \left(\frac{E}{E_0} \right)^{-\Gamma} \label{eq:pow} ,
\end{equation}
and a log-parabola
\begin{equation}
  \frac{dN}{dE} = N_0 \left(\frac{E}{E_b} \right)^{-(\Gamma + \beta \log(E/E_b))} \label{eq:logp},
\end{equation}
with $E_b=308$\,MeV fixed to the value reported in the 3FGL catalogue.

For the considered period between August 2015 and May 2017, \cta is detected with a Test Statistic \citep[TS,][]{mea96} of 163879, i.e. $\sim$405$\sigma$. The spectrum of \cta is significantly curved with a photon index of $\Gamma=2.068 \pm 0.008$ and a curvature index of $\beta=0.064 \pm 0.003$. The average flux is $F=(2.27 \pm 0.01) \times 10^{-6}$\,ph\,cm$^{-2}$\,s$^{-1}$.

To further study the activity of \cta at high energies, a light curve has been produced with a time-binning of 1\,day. Since on these time scales the preference of a log-parabola is not guaranteed, the spectrum has been modeled with a simple power-law in each time bin, leaving the photon index free to vary. The resulting light curve is shown in Fig.~\ref{fig:lc}(a).

From this data set, spectra were derived for two particular dates: MJD\,57670 and MJD\,57745, which are representative of the pre-flare state and the flare state around the maximum.
For MJD\,57670, \cta is detected with TS=161 ($\sim 12\sigma$), and the observed spectrum is well described by a power-law with $F=(1.19 \pm 0.23) \times 10^{-6}$\,ph\,cm$^{-2}$\,s$^{-1}$ and $\Gamma=2.08 \pm 0.14$. Testing a log-parabola only yielded a log-likelihood ratio 0.2 with respect to a power-law. In order to validate that the non-detection of curvature is independent of the detection significance, we derived a 10-day spectrum starting on MJD\,57670. Despite the increased significance of the source with TS=683 ($\sim 26\sigma$), the spectrum is still compatible with a power-law, since the log-parabola is only preferred with $0.95\sigma$.
For MJD\,57745, the detection level of \cta reaches TS=4558 ($\sim 67\sigma$), and the observed source spectrum is significantly curved, with a log-likelihood ratio of 9.9 for a log-parabolic spectrum with respect to a power-law. The corresponding spectrum results in $F=(1.10 \pm 0.05) \times 10^{-5}$\,ph\,cm$^{-2}$\,s$^{-1}$, $\Gamma=1.797 \pm 0.061$ and $\beta=0.077 \pm 0.025$.
The two 1-day spectra are shown in Fig.~\ref{fig:modspec} as the black and red butterfly, respectively. The displayed spectra have been corrected for absorption by the extragalactic background light (EBL) following the model of \cite{frv08}, which has, however, only a minor influence at the highest energies.

The change in spectral shape can be interpreted as a move of the peak energy during the flare towards higher energies. While the peak of the IC component cannot be determined before the flare (somewhere between 10\,keV and 100\,MeV), during the peak of the flare it is at about 3\,GeV. This points towards a significant hardening of the underlying particle distribution.

\subsection{X-ray analysis} \label{sec:xray}
The Swift Gamma-Ray Burst Mission  \citep[hereafter \textit{Swift},][]{gea04} is a multi-frequency space observatory which allows to monitor  targets in the optical, ultraviolet and X-ray energy bands. 
The X-ray Telescope \citep[XRT,][]{bea05} monitored \cta since 2005 in 137 pointing observations taken in the energy range of 0.3-10\,keV.
In this work, the lightcurve (Fig.~\ref{fig:lc}(b)) presents data collected between MJD 57668 and MJD 57821, which correspond to the ObsIDs of 00033509084-00033509120.

All data collected were analysed using version 6.20 of the HEASOFT package.\footnote{\url{http://heasarc.gsfc.nasa.gov/docs/software/lheasoft}}
The data were recalibrated using the standard procedure \verb|xrtpipeline|.
For the spectral fitting \verb|XSPEC| v.12.8.2 was used \citep{a96}.
All data were binned to have at least 30 counts per bin. 
Each observation has been fitted using the power-law  model, Eq. (\ref{eq:pow}), with the Galactic absorption value of $N_{H} = 4.76 \times 10^{20}$\,cm$^{-2}$ \citep{kea05} set as a frozen parameter.
In each observation we checked also if a broken power-law  model can result in a better description of the spectrum. 
According to reduced $\chi^2$ values, a simple power-law is the best model for all data in our set. 

The two observations presented in the global SED (Fig. \ref{fig:modspec} are described with the following spectral parameters:
$\Gamma_{57670} = 1.3 \pm 0.2$ and $N_{57670}= (1.17 \pm 0.16) \times 10^{-3}\,$cm$^{-2}$s$^{-1}$keV$^{-1}$ and 
$\Gamma_{57745} = 1.52 \pm 0.06$ and $N_{57745}= (3.93 \pm 0.18) \times 10^{-3}\,$cm$^{-2}$s$^{-1}$keV$^{-1}$.
The spectrum shown with black symbols corresponds to observations taken nearest to MJD 57670, which is data with ObsId 00033509084, while red symbols correspond to the observations taken nearest to MJD 57745, which is data with ObsId 00033509109. Apparently, only the normalization of the spectra changes.

\subsection{Optical/UV analysis} \label{sec:optical}
Simultanously with XRT, \cta was monitored with the UVOT instrument onboard \textit{Swift}.
The observations were taken in the UV and optical  bands with the central wavelengths of: UVW2 (188 nm), UVM2 (217 nm), UVW1 (251 nm), U (345 nm), B (439 nm), and V (544 nm). 
The instrumental magnitudes were calculated using the \verb|uvotsource| task including all photons from a circular region with radius 5''.
The background was determined from a circular region with a radius of 5'' near the source region that is not contaminated with signal from any nearby source. 
The optical and ultraviolet data points were corrected for dust absorption using the reddening $E(B-V)$ = 0.0612\,mag \citep{sf11} and the ratios of the extinction to reddening, $A_{\lambda} / E(B-V)$ \citep{gea06}. 

Further optical data in R- and B-band filters have been obtained with the Automatic Telescope for Optical Monitoring (ATOM), which is a $75\,$cm optical telescope located at the H.E.S.S. site in the Khomas Highland in Namibia \citep{hea04}. 
It regularly observes roughly 250 $\gamma$-ray emitters.

ATOM monitors \cta since 2008. 
During the visibility period presented in this paper, R-band monitoring lasted  from June 2016 until January 2017. 
Additional B-band observations were taken from October 2016 until December 2016.
Most of the high-flux period is covered by at least one B-band and several R-band measurements per night.
The data were analyzed using the fully automated ATOM Data Reduction and Analysis Software and have been manually quality checked.
The resulting flux was calculated via differential photometry using 5 custom-calibrated secondary standard stars in the same field-of-view.

Using measurements from a calm period between 2008 and 2011 the baseline flux of \cta can be established as $R = 16.90 \pm 0.02\,$mag.
An outburst in September 2012 reached $R = 14.6 \pm 0.1$mag before returning to previous levels. In late 2015, ATOM detected \cta at $R = 16.54 \pm 0.08\,$mag.
Beginning in mid 2016, \cta showed increasing activity with a first outburst in August reaching $R = 14.20 \pm 0.02\,$mag.
Towards the end of visibility \cta started to steadily brighten, culminating in $R = 10.96 \pm 0.05\,$mag on 29 December 2016 (MJD 57751). We find significant intra-night variability, similar to the results reported in \cite{bea17}.
Both R- and B-band lightcurves are shown in Fig.~\ref{fig:lc}(c).

We have confirmed that the color of the optical/UV spectra is constant in time, which implies that the peak of the synchrotron component does not move significantly from its initial, unknown position in the infrared towards bluer, optical frequencies. This has the unfortunate side-effect that we cannot determine the peak synchrotron energy during this flare. On the other hand, one can deduce that neither the maximum Lorentz factor of the electrons nor the magnetic field increase significantly.

\subsection{Flux evolution after March 2017} \label{sec:mwlafter}
Between mid January and late April the source is not visible for optical and X-ray observatories, since \cta is too close to the sun during these months. Hence, the downward trend visible in the \g-ray lightcurve could not be observed in any other band. Swift and ATOM resumed observations of \cta in late April. 

The optical flux was still highly variable between $R = 16\,$mag and $R = 13\,$mag while displaying a general trend of fainting. The behavior in the X-ray band was similar. In the \g-ray domain, fluxes became variable again in early April exhibiting day-long outbursts similar to the behavior before October 2016. Therefore, we conclude that the optical and X-ray activity at that time is unrelated to the \g-ray activity between October 2016 and March 2017 and of no concern for our modeling.

%
%
\section{Cloud ablation by the relativistic jet} \label{sec:theory}
\begin{figure*}[t]
\begin{minipage}{0.49\linewidth}
\centering \resizebox{\hsize}{!}
{\includegraphics{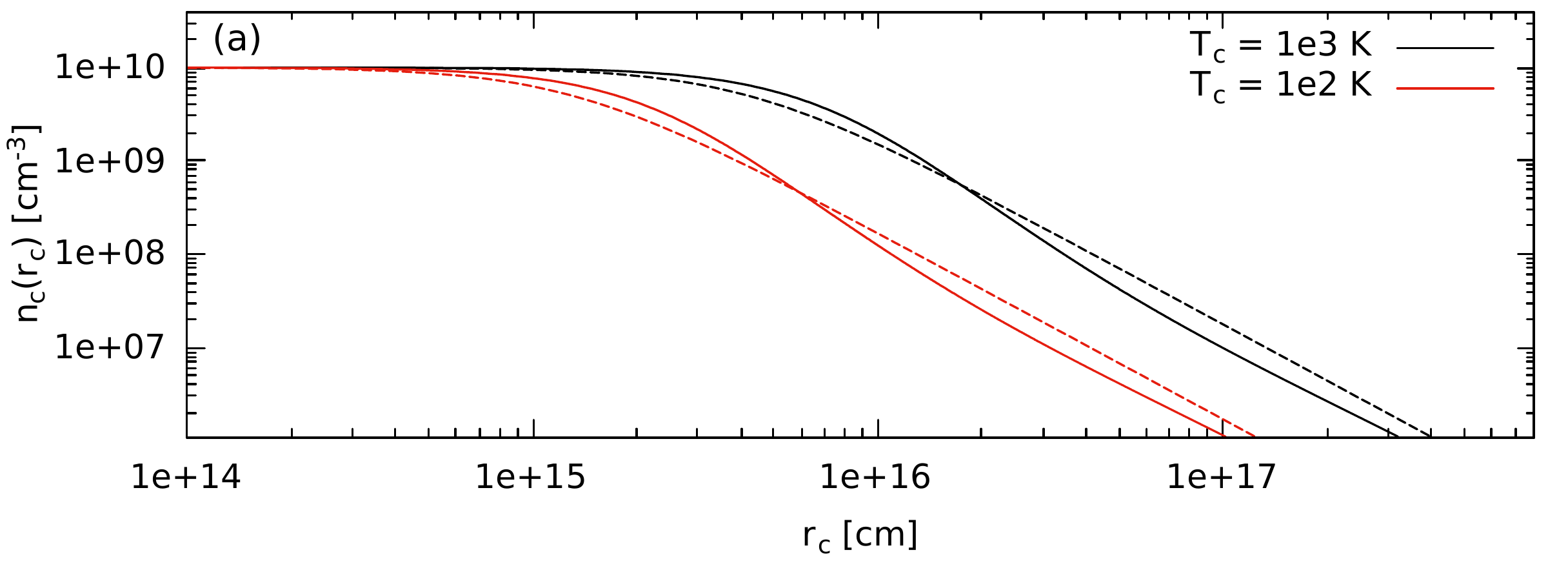}}
\end{minipage}
\hspace{\fill}
\begin{minipage}{0.49\linewidth}
\centering \resizebox{\hsize}{!}
{\includegraphics{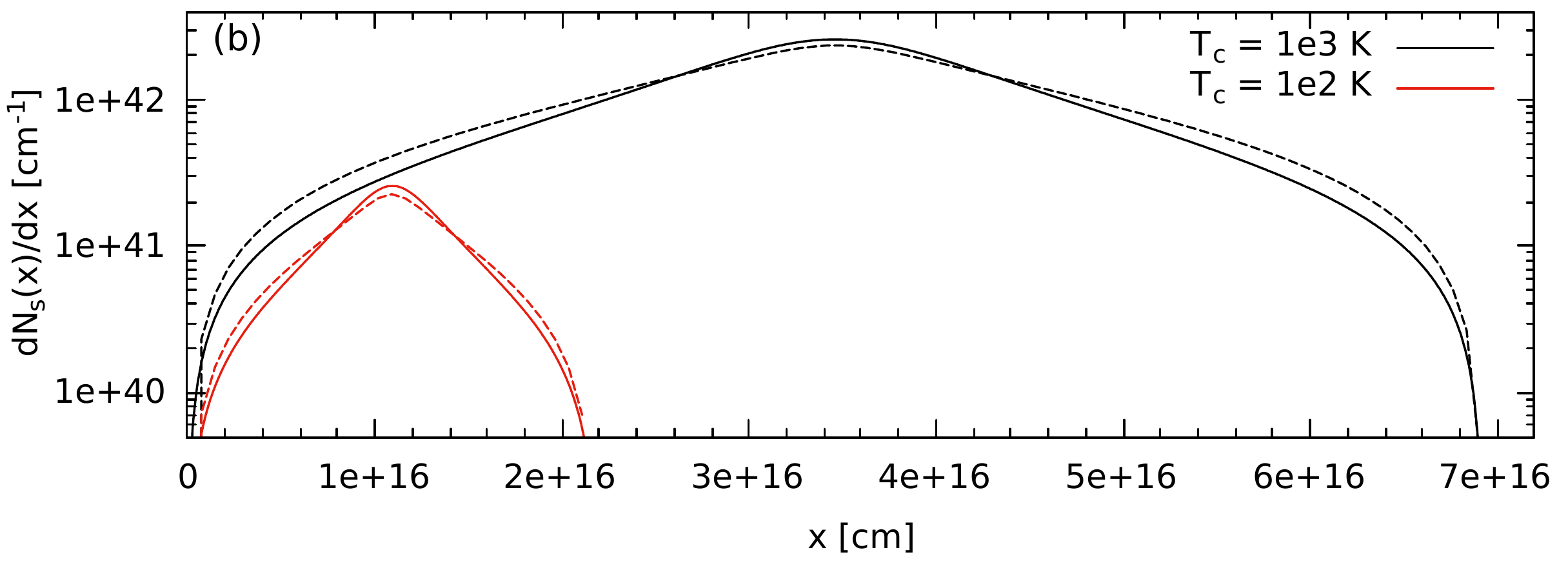}}
\end{minipage}
\caption{{\bf (a)} Numerical (solid) and approximate (dashed) solution for the cloud density distribution $n_c\p$, Eq. (\ref{eq:hydrostat2}), as a function of cloud radius $r_c\p$ for two values of the cloud temperature $T_c\p$ as labeled. The central density is set to $n_0\p = 10^{10}\,$cm$^{-3}$. {\bf (b)} Integration of the numerical solution (solid) and the analytical solution, Eq. (\ref{eq:ns}), (dashed) of the cloud density distribution as a function of slice position $x\p$ for two values of its temperature $T_c\p$ as labeled. Parameters as in (a). In both panels, we dropped the primes for clarity.}
\label{fig:cloud}
\end{figure*} 

The potential cause of a strong outburst is the accumulation of more matter than usual in a standing shock within the jet. If a gas cloud on its orbit around the black hole happens to penetrate the jet, it will be ablated and carried along by the jet. Hence, this is an efficient process for the jet to pick up a large amount of material and to cause prolonged jet activity, if the cloud is ablated at a steady pace while it enters the jet. Given the changing density within the cloud, the jet ablates different amounts of matter at a given time while the cloud penetrates the jet. This leads to a gradual increase and decrease of the lightcurve during the flare over the time scale the cloud is ablated. Density fluctuations within the cloud and instabilities during the process might lead to a more chaotic ablation, which could result in strong and fast fluctuations related to the size of these fluctuations on top of the longer trend. In the following, we will concentrate on the long-term trend, and discuss the influence of density fluctuations elsewhere.

The situation is that a spherical cloud approaches the jet with orbital speed around the central black hole
\begin{align}
 v_c\p = \sqrt{G M_{\rm bh}/z\p} \label{eq:vc} ,
\end{align}
where $G$ is the gravitational constant, and $z\p$ the distance between the cloud and the black hole. The radius of the cloud can be derived from the rising time $t_f\p$ (that is from the beginning to the peak) of the event:
\begin{align}
 R_c\p = t_f\p v_c\p = \frac{t_f^{\rm obs} v_c\p}{(1+\zred)}  \label{eq:Rc} .
\end{align}
Apart from the redshift correction, the frame of the cloud and the observer's frame are identical, since the motion of the cloud is non-relativistic. Hence, the observed duration of the flare is indeed the same as the cloud penetration time.

The number of particles in the cloud follows from the increase in particles in the jet, under the assumption that the cloud is fully ablated. We can calculate the particles in the cloud, if we take the difference of particles at the peak and at the beginning of the flare. This includes the simplifying assumption that the cloud contains a pure hydrogen plasma. Within the emission region of the jet the density of electrons (and possibly positrons) is $n_{j,e}$. The electron charge is balanced by a fraction $a\leq 1$ of protons, depending on the number of positrons in the jet. The total density of particles in the jet is $n_{j} = (1+a)n_{\rm j,e}$. The number of particles in the emission region obviously is $N_j = \frac{4}{3}\pi R_j^3 n_j$, which is an invariant quantity. $R_j$ is the radius of the jet. Hence, the number of particles in the cloud equals the difference of particle number in the jet at the peak of the event to the beginning of the event:
\begin{align}
 N_c\p &= N_c = 2(N_{\rm j,max} - N_{\rm j,min}) \nonumber \\
 &= \frac{8\pi}{3} R_j^3 (1+a) (U-1) n_{j,e,{\rm min}} \label{eq:Nc} .
\end{align}
The factor 2 takes into account that the maximum of the event takes place when the center of the cloud is ablated and the second half of the cloud is still to be ablated. $U = n_{j,e,{\rm max}}/n_{j,e,{\rm min}}$ marks the ratio of the electron densities at the peak and the beginning of the flare. Naturally, in the cloud $a=1$. Hence, the addition of cloud material into the jet should raise the value of $a$ in the jet emission region. For ease of computation, we neglect this effect here. For an initial jet plasma with $a\lesssim 1$, the influence is negligible.

The jet ablates the cloud due to its ram pressure, which is in the black hole frame
\begin{align}
p_{\rm ram}\p &= (\Gamma_j-1) n_{j,e,{\rm min}}\p \bar{\gamma}_e m_ec^2 + (\Gamma_j-1) n_{j,p,{\rm min}}\p m_pc^2 \nonumber \\
&= \Gamma_j (\Gamma_j-1) \bar{\gamma}_e m_ec^2 \left( 1+a\frac{m_p}{\bar{\gamma}_e m_e} \right) n_{j,e,{\rm min}} \label{eq:pram} ,
\end{align}
introducing the bulk Lorentz factor $\Gamma_j$ of the jet, and the speed of light $c$. Not surprisingly, for fractions of protons $1~\geq~a~>~\bar{\gamma}_e~m_e~/~m_p$, with the average electron Lorentz factor $\bar{\gamma}_e$, the electron mass $m_e$, and the proton mass $m_p$, the ram pressure is dominated by protons. Since the ram pressure of the jet is provided by particles already present in the jet, it remains constant throughout the flare and can be reconstructed by pre-flare parameters.

The gravitational pressure that keeps the cloud together, is
\begin{align}
 p_g\p(r_c) = \frac{F_g\p(r_c\p)}{A_H} = \frac{GM_c(r_c\p)m_H}{\pi r_H^2 r_c^{\prime 2}} \label{eq:pg} ,
\end{align}
where $A_H=\pi r_H^2\sim 8.8\times 10^{-17}\,$cm$^2$ is the cross-section and $m_H\sim m_p$ is the mass of a hydrogen atom, which constitutes the bulk of the particles in the cloud. $M_c(r_c\p)$ is the enclosed mass at cloud radius $r_c\p$. 

The cloud will be ablated, if $p_{\rm ram}\p>p_g\p$. Hence, with Eqs. (\ref{eq:pram}) and (\ref{eq:pg}), and a slight redistribution, we can construct a lower limit on the initial jet electron density:
\begin{align}
 n_{j,e,{\rm min}} > \frac{Gm_H}{\pi r_H^2 m_ec^2} \frac{M_c(r_c\p)}{\Gamma_j (\Gamma_j-1) \bar{\gamma}_e \left( 1+a\frac{m_p}{\bar{\gamma}_e m_e} \right) r_c^{\prime 2}} \label{eq:njmin}
\end{align}
In order to get an estimate on the required jet electron density, we chose the outer layer of the cloud $r_c\p=R_c\p$ as an example. Approximating $\bar{\gamma}_e\ll m_p/m_e$ and $m_H\sim m_p$, we find
\begin{align}
 n_{j,e,{\rm min}} &\gtrsim 2.8\times 10^{-12} \left( \frac{a}{0.1} \right)^{-1} \left( \frac{\Gamma_j}{10} \right)^{-1} \left( \frac{\Gamma_j -1}{9} \right)^{-1} \nonumber \\
 &\quad \times \left( \frac{M_c}{0.01 M_{\odot}} \right) \left( \frac{R_c\p}{10^{15}\,\mbox{cm}} \right)^{-2} \,\mbox{cm}^{-3} \label{eq:njminEST} .
\end{align}
Obviously, the cloud cannot withstand destruction. Even a solar-like star with much higher surface gravity could be stripped off its outer layers while penetrating the jet, which typically exhibits electron densities exceeding $10^{-2}\,$cm$^{-3}$. However, this estimate might not hold for the inner, dense core of a star.

Given that the cloud penetrates the jet gradually, the number of particles injected into the jet changes over time. In order to calculate the correct injection term, the density distribution of the cloud $n_c\p(r_c\p)$ must be known. We consider a profile based on hydrostatic equilibrium. The simplest ansatz would be to assume that the cloud consists of isothermal ideal gas with temperature $T_c\p$, so that the thermal pressure $p_T\p = \rho_c\p k_B T_c\p / m_p$, where $\rho_c\p = m_p n_c\p$ is the cloud's mass density, and $k_B$ is the Boltzmann constant. In this case, the equation of hydrostatic equilibrium reads
\begin{align}
 \frac{k_B T_c\p}{m_p} \frac{\td{\rho_c\p(r_c\p)}}{\td{r_c\p}} &= - g(r_c\p) \, \rho_c\p(r_c\p) \nonumber \\
 &= - 4 \pi \frac{G \rho_c\p(r_c\p)}{r_c^{\prime 2}} \int\limits_{0}^{r_c\p} \td{\tilde{r}} \tilde{r}^2 \rho_c\p(\tilde{r}) \label{eq:hydrostat} .
\end{align}

With the definition $\tau\p \equiv k_B T_c\p / (4 \pi \, m_p \, G)$, Eq. (\ref{eq:hydrostat}) reduces to
\begin{align}
 \tau\p \, \frac{\td{}}{\td{r_c\p}} \left( \frac{r_c^{\prime 2}}{\rho\p} \, \frac{\td{\rho\p}}{\td{r_c\p}} \right) = - \rho\p \, r_c^{\prime 2} \label{eq:hydrostat2} .
\end{align}
The numerical solution to this nonlinear differential equation is plotted in Fig. \ref{fig:cloud}(a) for two values of $T_c\p$. As is also shown in that plot, the numerical solution is well approximated by:
\begin{align}
 n_c\p(r_c\p) = \frac{n_0\p}{1+\left( r_c\p / r_0\p \right)^2} \label{eq:ncloud} .
\end{align}
The normalization $n_0\p$ can be determined by integrating Eq. (\ref{eq:ncloud}) and equating it to Eq. (\ref{eq:Nc}), and 
\begin{align}
 r_0\p = \sqrt{\frac{3\tau\p}{m_p n_0\p}} \label{eq:r0p} .  
\end{align}

Naturally, the density drops to zero for $r_c\p\rightarrow \infty$. In order to make progress, we approximate the cloud as a sphere with outer boundary $R_c\p>r_0\p$ and set $n_c\p(r_c\p\geq R_c\p)=0$. Once the cloud hits the jet, it is ablated slice-by-slice beginning with a low particle-number region at the front, through the dense central region, and ending again at a low-density region at the rear side. Therefore, we define all quantities of the cloud as a function of $x\p$, the slice position with respect to the outer edge of the cloud that first touches the jet. That is, $x\p=0$ where the cloud first touches the jet, $x\p=R_c\p$ is the cloud's center, and $x\p=2R_c\p$ marks the rear side of the cloud. With the speed of the cloud, it can be written as $x\p = v_c\p t\p$, where $t\p$ is the time that has passed since first contact in the AGN frame.

The number of particles ablated in each slice is the integral over the density $n_c\p(r_c\p)$ with respect to the slice volume. In the case of a sphere, the volume of a slice between positions $x\p$ and $x\p+\td{x\p}$ is \citep{zs13}
\begin{align}
 \td{V_s\p(x\p)} = \td{x\p} \int \td{A_s\p(x\p)} = \pi \left( 2R_c\p x\p-x^{\prime 2} \right) \td{x\p} \label{eq:Vslice} ,
\end{align}
where $A_s\p(x)$ is the cross-section of a slice, and $\td{x\p}$ its width.
The particle number in each slice then becomes
\begin{align}
 \td{N_s\p(x\p)} = \td{x\p} \int n_c\p(r_c\p) \td{A_s\p(x\p)} \label{eq:ns1} .
\end{align}
Writing the integral in cylindrical coordinates with $r_c\p(x\p) = \sqrt{\omega^2 + (R_c\p-x\p)^2}$, and $\omega_c\p(x\p) = \sqrt{2R_c\p x\p-x^{\prime 2}}$, Eq. (\ref{eq:ns1}) becomes
\begin{align}
 \td{N_s\p(x\p)} = 2\pi\td{x\p} \intl_0^{\omega_c\p(x\p)} n_c\p(r_c\p(\omega))\, \omega \td{\omega} \label{eq:ns2} .
\end{align}
Inserting Eq. (\ref{eq:ncloud}) in Eq. (\ref{eq:ns2}), the integral can be easily solved, giving
\begin{align}
 \td{N_s\p(x\p)} = \pi\td{x\p}r_0^{\prime 2}n_0\p \logb{\frac{r_0^{\prime 2} + R_c^{\prime 2}}{r_0^{\prime 2}+(R_c\p-x\p)^2}} \label{eq:ns} .
\end{align}
This function is shown in Fig. \ref{fig:cloud}(b) for two cases of $T_c$ along with an integration of the numerical solution of Eq. (\ref{eq:hydrostat2}). The analytical approximation and the exact result match nicely.

The injection of particles in the jet, which get dragged along and cause the flare at a shock somewhere downstream, can then be described by
\begin{align}
 Q_{\rm inj}(t) \propto \logb{\frac{r_0^{\prime 2} + R_c^{\prime 2}}{r_0^{\prime 2}+(R_c\p-x\p)^2}} \DF{t-\frac{x\p}{v_c\p}} \label{eq:injection} .
\end{align}
Here, $\DF{q}$ is Dirac's $\delta$-function, which describes the slice-by-slice ablation in time.

We stress that the entire mass of the cloud is not added to the jet at once, but gradually over about 4 months in the observer's frame. Hence, the impact of the added mass on the jet's bulk Lorentz factor at any given time is minor compared to a case where the entire cloud mass would be added at once. In the following, we assume a constant jet bulk Lorentz factor.

%
\section{Modeling} \label{sec:mod}
%
%
%
\begin{figure}[t]
\centering 
\includegraphics[width=0.48\textwidth]{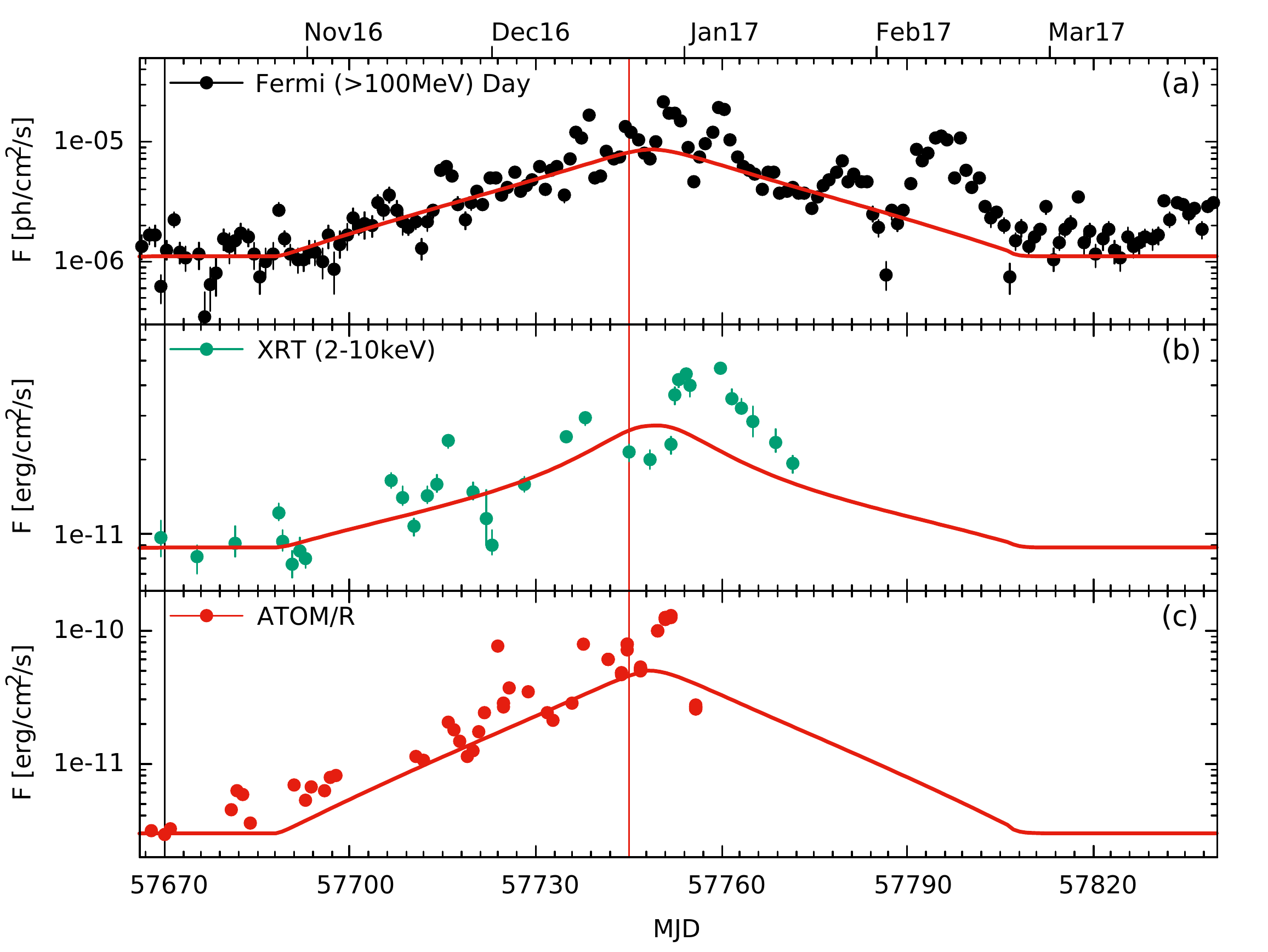}
\caption{Lightcurves of {\bf (a)} \fermi data, {\bf (b)} {\it Swift}-XRT data, and {\bf (c)} ATOM/R data. The thick red lines are the modeling result, while the vertical thin black and red lines mark the dates, where the spectra have been extracted. Note the logarithmic scaling of the y axis.}
\label{fig:modlc-log}
\end{figure} 
\begin{figure}[t]
\centering
\includegraphics[width=0.48\textwidth]{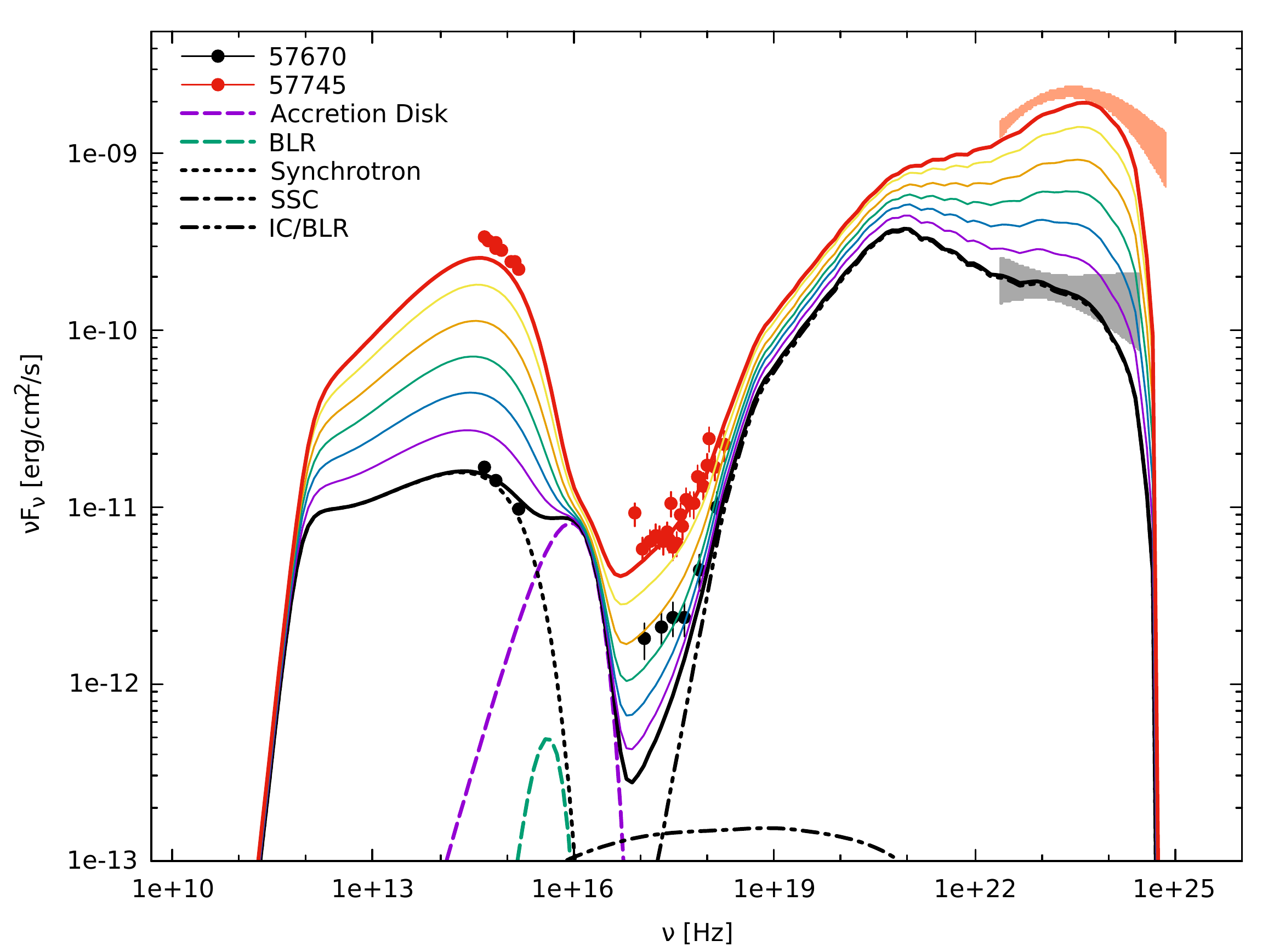}
\caption{Two representative spectra of CTA~102 during the flare: MJD 57670 (black symbols and butterfly), and MJD 57745 (red symbols and butterfly). The \fermi\ spectra have been corrected for EBL absorption using the model of \cite{frv08}. The thick black and red solid lines show model spectra for the beginning and the peak of the flare, while the thin solid lines (magenta, green, orange, yellow, blue) show the evolution of the model spectrum in roughly 10-day steps towards the maximum. The other lines give example curves of the composition of the spectrum: accretion disk (dashed magenta), BLR (dashed green), synchrotron (dotted black), SSC (dash-dotted black), IC/BLR (dash-double-dotted black).}
\label{fig:modspec}
\end{figure} 
\begin{figure}[t]
\centering
\includegraphics[width=0.48\textwidth]{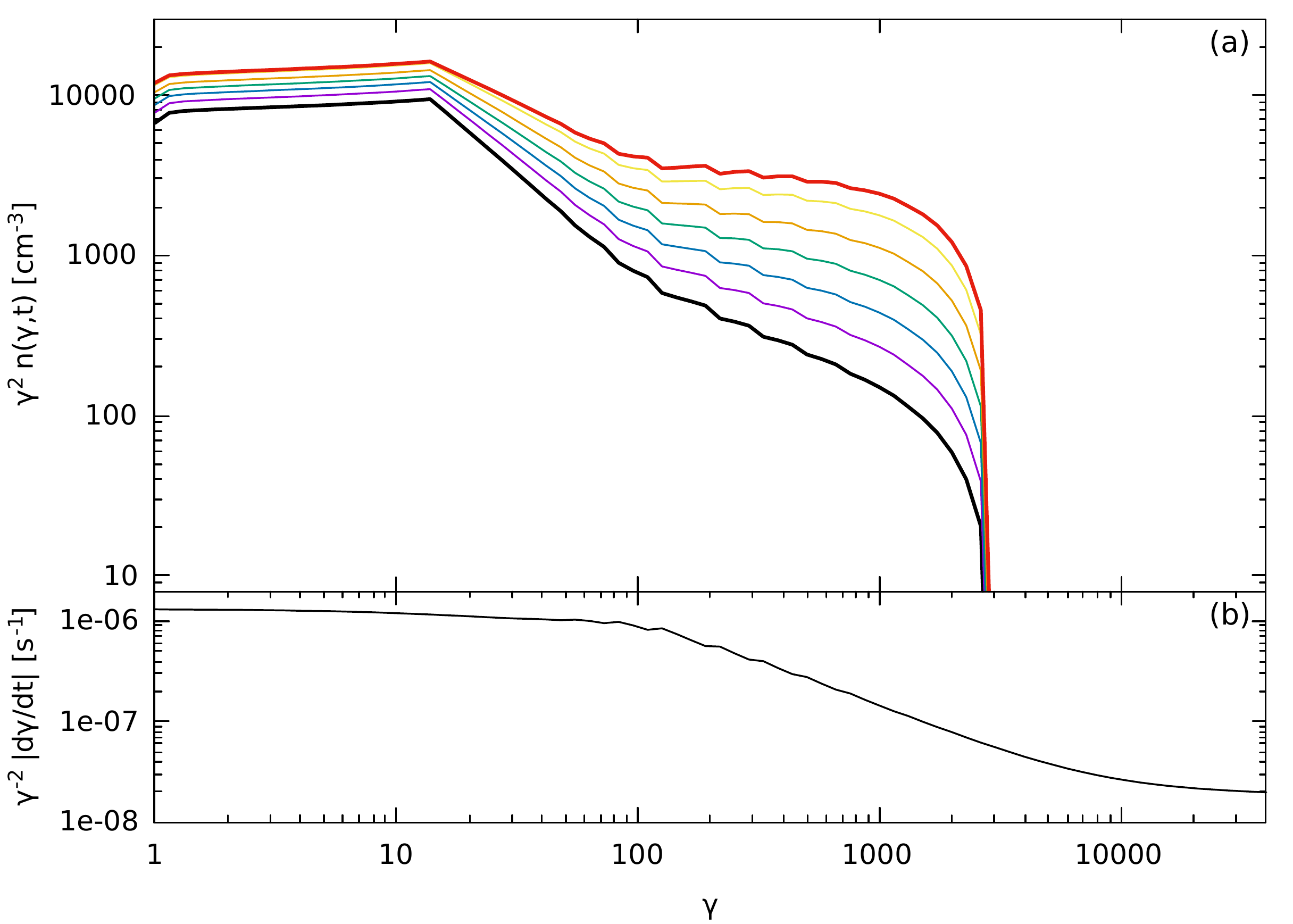}
\caption{{\bf (a)} Electron distribution function $\gamma^2 n(\gamma,t)$ as a function of the electron Lorentz factor $\gamma$ for the same time steps as in Fig. \ref{fig:modspec}. {\bf (b)} Electron cooling term $\gamma^{-2} |\dot{\gamma}|$ as a function of the electron Lorentz factor $\gamma$.}
\label{fig:modpart}
\end{figure} 
\begin{table}
\caption{Model parameter description, symbol and value. Values below the horizontal line mark parameters for the induced variability.}
\begin{tabular}{lcc}
Definition				& Symbol 			& Value \\
\hline
Emission region distance to black hole	& $z\p$				& $6.5\times 10^{17}\,$cm \\ 
Doppler factor of emission region	& $\delta_j$			& $35\,$ \\ 
Emission region radius			& $R_j$				& $2.5\times 10^{16}\,$cm \\ 
Magnetic field of emission region	& $B_j$				& $3.7\,$G \\ 
Electron injection luminosity		& $L_{j,e,{\rm inj}}$		& $2.2\times 10^{43}\,$erg/s \\ 
Minimum electron Lorentz factor		& $\gamma_{\rm min}$		& $1.3\times 10^1\,$ \\ 
Maximum electron Lorentz factor		& $\gamma_{\rm max}$		& $3.0\times 10^3\,$ \\ 
Electron spectral index			& $s$				& $2.4\,$ \\ 
Escape time scaling			& $\eta_{\rm esc}$		& $10.0\,$ \\ 
Acceleration to escape time ratio	& $\eta_{\rm acc}$		& $1.0\,$ \\ 
Effective temperature of the BLR	& $T_{\rm BLR}\p$		& $5.0\times 10^4\,$K \\
\hline
Electron luminosity variation		& $\Delta L_{j,e,{\rm inj}}$	& $1.75\times 10^{43}\,$erg/s \\ 
Electron spectral index variation	& $\Delta s$			& $-0.6\,$ \\ 
Time between onset and peak of flare	& $t_f^{\rm obs}$		& $60\,$d \\
Cloud scale height 			& $r_0\p$			& $1.6\times 10^{14}\,$cm \\
\end{tabular}
\label{tab:inputcom}
\end{table}
In order to model the long-term trend of the \cta flare, we use the code by \cite{db14} and adapt it slightly to accommodate the variability induced by the cloud ablation as discussed in section \ref{sec:theory}. The code calculates the electron distribution and photon emission spectra in the comoving frame of the emission region, and subsequently transforms it to the observer's frame taking into account the Doppler factor $\delta_j$, which we assume here to be equal to the bulk Lorentz factor $\Gamma_j$, and the redshift $z_{\rm red}$. The electron distribution function $n_e(\gamma,t)$ is calculated with a Fokker-Planck-type differential equation that takes into account injection, stochastic acceleration, cooling and escape.

The injection electron distribution is of the form
\begin{align}
 Q(\gamma,t) = Q_0(t) \gamma^{-s(t)} \HF{\gamma; \gamma_{\rm min}(t), \gamma_{\rm max}(t)} \label{eq:elinj},
\end{align}
where $\gamma$ is the electron Lorentz factor, $s$ the electron spectral index, and $\HF{\gamma; \gamma_{\rm min}, \gamma_{\rm max}}$ denotes Heaviside's step function with $H=1$ for $\gamma_{\rm min}\leq \gamma\leq \gamma_{\rm max}$ and $H=0$ otherwise. The injection normalization is derived from input parameters as
\begin{align}
 Q_0(t) = \frac{L_{j,e,{\rm inj}}(t)}{V_j m_ec^2}
 \begin{cases}
  \frac{2-s(t)}{\gamma_{\rm max}^{2-s(t)} - \gamma_{\rm min}^{2-s(t)}} & \mbox{if} \,\, s\neq 2 \\
  \left( \ln{\frac{\gamma_{\rm max}}{\gamma_{\rm min}}} \right)^{-1} & \mbox{if} \,\, s=2
 \end{cases} \label{eq:Q0el} ,
\end{align}
with the electron injection luminosity $L_{j,e,{\rm inj}}$, and the comoving volume $V_j=\frac{4}{3}\pi R_j^3$ of the emission region. Since the input parameters can be time-dependent, the injection distribution might change in every time step.

The acceleration and escape terms are parameterized independent of energy. The escape time scale is defined by $t_{\rm esc} = \eta_{\rm esc} R/c$, namely a multiple $\eta_{\rm esc}$ of the lightcrossing time scale. The acceleration time scale in turn is defined as a multiple $\eta_{\rm acc}$ of the escape time scale: $t_{\rm acc} = \eta_{\rm acc} t_{\rm esc}$.

The cooling term takes into account all radiative processes, namely synchrotron radiation in a randomly oriented magnetic field $B_j$, SSC, and IC emission on potential external fields, such as the accretion disk, the BLR or a dusty torus. The IC process takes into account the full Klein-Nishina cross section. The accretion disk spectrum is assumed to be of the \cite{ss73} type, which basically depends on the mass of the central black hole and the Eddington ratio $\eta_{\rm Edd}$ of the disk luminosity $L_{\rm disk}\p = \eta_{\rm Edd} L_{\rm Edd}\p$. The BLR spectrum is assumed to be a black-body spectrum of effective temperature $T_{\rm BLR}\p$ normalized to the measured BLR luminosity. The size of the BLR is important to calculate the BLR energy density and potential absorption of $\gamma$ rays from the emission region. Similar definitions are possible for the dusty torus, but we neglect that photon field here due to a lack of observational evidence.

An implicit Crank-Nichelson scheme is used to solve the Fokker-Planck equation. With the solution for $n_{e}(\gamma,t)$ the radiation spectra are derived, which consider internal absorption through synchrotron-self absorption, and external absorption of $\gamma$ rays through the external soft photon fields.

Before starting to model the lightcurve, we first derived two exemplary spectra of the low state before the flare in October and of the high state in late December in order to derive the baseline parameter sets that needed to be matched at these two states. The dates are MJD 57670 and MJD 57745, and are marked by black and red vertical lines in the lightcurves of Fig.~\ref{fig:modlc-log}, respectively. We have chosen these dates, since in addition of being representative of the respective flux levels, the data taken in all bands is contemporaneous. The spectra are shown in Fig.~\ref{fig:modspec}. Most obvious are the significant flux changes between the two states and the change in peak energy of the IC component. The parameters of the fit to the low state are given in Tab.~\ref{tab:inputcom}.

A few of these parameters are constrained by observations. 
The size of the emission region modulo Doppler factor $R_j/\delta_j$ is constrained by the variability time scale in our data as $R_j \lesssim \Delta t^{\rm obs} c \delta_j /(1+z_{\rm red})$. Due to the measured optical intranight variability, the emission region must be smaller than a lightday in the observer's frame corresponding to less than $4.5\times 10^{16}\,$cm for a Doppler factor of $\delta_j=35$. The chosen value of $R_j=2.5\times 10^{16}\,$cm is a compromise between the aforementioned limit and the necessity of a rather large emission region in order to keep the SSC emission low. The latter would quickly overproduce the X-ray flux for smaller source radii especially during the variable period.

The magnetic field $B_j$ is constrained from the Compton dominance parameter $W$, which is defined as the ratio of the peak fluxes of the two spectral components. The peak fluxes are directly proportional to the underlying energy densities, namely the magnetic energy density and the energy density in the BLR photon field transformed to the comoving frame. Hence, $W = 4\Gamma_j^2u_{\rm BLR}\p / 3u_{\rm B}$. Solving for $B_j$, one obtains
\begin{align}
 B_j &= \sqrt{\frac{8\Gamma_j^2 L_{\rm BLR}\p}{3c R_{\rm BLR}^{\prime 2}W}} \nonumber \\
 &= 2.9\,\left( \frac{\Gamma_j}{10} \right)\,\left( \frac{W}{10} \right)^{-1/2} \left( \frac{L_{\rm BLR}\p}{4.14\times 10^{45}\,\mbox{erg/s}} \right)^{1/2} \nonumber \\
 &\quad \times \left( \frac{R_{\rm BLR}\p}{6.7\times 10^{17}\,\mbox{cm}} \right)^{-1}  \, \mbox{G} \label{eq:Bconst}.
\end{align}
Unfortunately, the peak fluxes of both the synchrotron and the inverse Compton component are not well defined in the low state. Hence, $W$ is not particularly well constrained, and values of at least 10 are plausible. We follow the standard assumption of the one-zone model that the magnetic field is tangled. While this is a simplification, since one expects an ordered guide magnetic field in the jet, we have no observational constraints at hand that could constrain the geometry of the magnetic field during this particular event. \cite{lea16} modeled their polarimetry data of the 2012 flare assuming a helical magnetic field and a helical motion of the emission region, which is different from our model. 

The maximum electron Lorentz factor $\gamma_{\rm max}$ is constrained by the soft optical synchrotron spectrum. While a soft electron distribution could also account for the soft synchrotron spectrum, this would be inconsistent with the harder (than the optical spectrum) \g-ray spectrum. Hence, the soft optical spectrum can be interpreted as an exponential cut-off induced by a maximum electron Lorentz factor significantly below $10^4$. The minimum electron Lorentz factor $\gamma_{\rm min}$ is not constrained by the observations, but has been chosen in such a way that the IC/BLR spectrum fits well the hard X-ray spectrum. In principle, the \g-ray spectrum could be used to constrain the spectral index $s$. However, as one can see in Fig. \ref{fig:modpart}, the Klein-Nishina effect in the cooling changes the particle spectrum considerably at the particle energies that correspond to the \g-ray spectrum probed by \fermi\ (see also the discussion below). Hence, the standard relations between photon spectra and (un)cooled particle distributions do not work, and we chose the spectral index to match well the \fermi\ spectrum.

The observed luminosity of $L^{\rm obs}\sim 10^{48}\,$erg/s of the ground state is at the high end of FSRQ luminosities \citep[e.g., ][]{gea98}. In order to reduce the required particle energy densities, we chose a relatively high Doppler factor of $\delta_j=35$. However, observations of the MOJAVE program revealed radio knots moving with apparent speeds of $\sim 18c$ \citep{leaM16}, which permit Doppler and Lorentz factors in the chosen order of magnitude. 

While there is no observational constraint on the distance of the emission region from the black hole $z\p$, it is chosen in such a way that the emission region is immersed in BLR photons, but yet the attenuation of \g\ rays by the BLR photons is minimal. Closer to the black hole the attenuation would start to become important even in the HE domain resulting in softer spectra and a much poorer fit. A greater distance from the black hole would result in an inefficient IC/BLR process. Hence, the emission region should be located around the outer edge of the BLR. 

The results are insensitive to the escape time scaling $\eta_{\rm esc}$ and the acceleration to escape time ratio $\eta_{\rm acc}$, since the strong cooling (see Fig. \ref{fig:modpart}) dominates over escape and acceleration for all energies. The effective temperature of the BLR $T_{\rm BLR}\p$ is also not constrained by observations, but it impacts the onset of the Klein-Nishina domain in the inverse Compton process. The chosen value implies that the Klein-Nishina domain already sets in for electron Lorentz factors of $\sim 100$. Lower values of $T_{\rm BLR}\p$ would increase the electron turn-over energy slightly.

We note that the chosen parameter set is not unique, and other parameter sets might give equivalent results. However, the precise parameters are not important for the evolution of the event, which is the main concern of this paper. 

In order to model the evolution of the flare, we varied the electron injection luminosity following Eq. (\ref{eq:injection}) as
\begin{align}
 L_{j,e,{\rm inj}}(t) = L_{j,e,{\rm inj}} + \Delta L_{j,e,{\rm inj}}  \logb{\frac{t_0^2+t_f^2}{t_0^2+(t_f-t)^2}} \label{eq:Let} ,
\end{align}
where all parameters are considered in the comoving frame, implying $t_f = \delta_j t_f^{\rm obs}/(1+z_{\rm red})$, and the time scale $t_0 = \delta_j v_c\p r_0\p$ is related to the cloud's scale height $r_0\p$. There is no observational constraint on the latter value. We know that the cloud's scale height $r_0\p$ must be smaller than the cloud's radius $R_c\p$. We have tested a few values and found that the value related to the scale height given in Tab. \ref{tab:inputcom} gives the best fit. A larger scale height than the one used underpredicts the fluxes, while a smaller scale height produces a narrow peak, which is also inconsistent with the observations. 

In order to account for the changing peak energy of the IC component, we also change the electron spectral index. Due to lack of constraints, we assume a linear change as
\begin{align}
 s(t) = s + \Delta s \, \frac{t_f-|t_f-t|}{t_f} \label{eq:st} .
\end{align}
With $\Delta s$ being negative (see Tab.~\ref{tab:inputcom}), the injection spectrum hardens until the maximum of the flare and subsequently returns to the pre-flare value. We further assume that the bulk Lorentz factor of the emission region is constant.

The resulting model spectra are shown in Fig.~\ref{fig:modspec}. The fit of the pre-flare and high-state spectra (black and red curves) is quite good, taking into account that we do not aim for a precise fit. The colored spectra show the evolution of the spectrum from the low state towards the maximum in roughly 10-day intervals. The lack of evidence of a broken power-law spectrum in the X-ray domain, gives us confidence that the seemingly poor fit at low X-ray energies is not a big concern. The upturn of the model curves around $100\,$MeV is due to a change in the cooling behavior at these and higher energies, as shown in Fig. \ref{fig:modpart}(b). At low energies, the cooling is dominated by the IC/BLR process, but reduces for electron Lorentz factors $\gamma>100$ due to the Klein-Nishina effect. This hardens the electron distribution, as can be seen in Fig. \ref{fig:modpart}(a), where we show the underlying particle distribution of each photon spectrum of Fig. \ref{fig:modspec}. For electron Lorentz factors $\gamma>10^4$, synchrotron cooling becomes dominant, which is however unimportant for the present study, since we do not consider electrons with these energies. Since neither the BLR nor the magnetic field are assumed to vary, the electron cooling term is constant in time.
The small wiggles in the $\gamma$-ray spectra, the particle distributions, and the cooling term are due to numerical inaccuracies.

The resulting model lightcurves are shown as thick red lines in Fig.~\ref{fig:modlc-log}. We present the lightcurves with a logarithmic y-axis, in order to highlight the significant change in flux and the details of the theoretical lightcurve evolution. We model the $\gamma$-ray, X-ray and optical R-band. We neglect the other optical/UV bands, since the R-band is the most detailed synchrotron lightcurve, and the constant color implies that the behavior in the other bands is very similar. The model of the general evolution of the flare is very good in all energy bands.

%
\section{Discussion} \label{sec:dis}
The modeling gives a good representation of the overall flare profile. We can safely conclude that the long-term activity of \cta is consistent with the addition of a large amount of mass to the jet over a time period of a few months. We have modeled this by the penetration of the jet by a gas cloud, for which we only made the assumption of being in hydrostatic equilibrium. Below, we will discuss the potential origin of the cloud.

Given that we use the IC/BLR process to model the high-energy component of the spectrum, the cloud-jet interaction should take place within the BLR. We set the emission region close to the outer edge of the BLR, allowing the IC/BLR process to operate, while the absorption at \g rays is kept low. Hence, the cloud could originate from the BLR itself.

From the above modeling we can deduce that the electron density at the beginning of the flare is $n_{j,e,{\rm min}}=2.32\times 10^{4}\,$cm$^{-3}$, which rises to $n_{j,e,{\rm max}}=4.0\times 10^{4}\,$cm$^{-3}$ at the peak of the flare. With Eq.~(\ref{eq:Nc}), $a=0.1$, and the parameters given in Tab.~\ref{tab:inputcom}, we can calculate the number of particles in the cloud to $N_c\p = 2.34\times 10^{54}$ or a mass of $M_c = N_c\p m_p = 3.9\times 10^{30}\,\mbox{g}\sim 0.1\%\,M_{\odot}$. The speed of the cloud is, Eq. (\ref{eq:vc}), $v_c\p = 5.12\times 10^{8}\,$cm/s, and its radius, Eq. (\ref{eq:Rc}), $R_c\p = 1.3\times 10^{15}\,$cm. The average particle density in the cloud is, thus, $n_c\p = 2.54\times 10^{8}\,$cm$^{-3}$. The scale height of the cloud is $r_0\p=1.6\times 10^{14}\,$cm. This value along with Eq. (\ref{eq:Nc}), Eq. (\ref{eq:r0p}), and an integration of Eq. (\ref{eq:ncloud}), implies a temperature  
\begin{align}
 T_c\p &= \frac{Gm_p^2 N_c\p}{6k_B r_0\p \left[ (R_c\p/r_0\p) - \arctan{(R_c\p/r_0\p)} \right]} \nonumber \\
 &\sim 0.5\,\mbox{K} \label{eq:Temp} .
\end{align}

This temperature is clearly too low, since a gas cloud cannot become colder than the cosmic microwave background. Additionally, standard parameters for BLR clouds suggest a radius of $\sim 10^{13}\,$cm and an average density of $10^{9-11}\,$cm$^{-3}$ \citep{dea99,p06}. Hence, the size of our model cloud is too large, while the density is too low. However, all these parameters have been derived under the assumption that the entire cloud is devoured by the jet, and this does not include potential higher density regions responsible for the fast, but bright flares on top of the long-term trend. These higher density regions would exhibit higher temperatures, likely raising the temperature of the entire cloud.
Additionally, the collision of the cloud with the jet might induce a shock wave running through the former \citep[e.g., ][]{pfb02,abr09}, which could lead to ejection of cloud material away from the interaction site. Then our estimate is only a lower limit on the matter content of the cloud, and the particle number could be significantly higher, and hence the temperature. Furthermore, we have assumed that the hydrostatic equilibrium is solely mediated by an isothermal gas. If the cloud contains a significant magnetic field, it will stabilize the cloud even if the temperature is exceeding the isothermal temperature derived above.

While these considerations could lead to a density and temperature of the cloud that more closely resemble parameters of BLR clouds, it does not influence our estimate of the size $R_c\p$, which solely depends on the speed of the cloud. Since we assumed Keplerian motion of the cloud, the size of the cloud depends inversely on the square-root of the distance from the black hole.
Hence, the size of the cloud can be reduced, if the ablation takes place further away from the black hole. While this could bring the size closer to the BLR cloud parameters, a BLR cloud can be excluded, since our model is already placed close to the outer edge of the BLR. In such a case, the high-energy component cannot be due to IC/BLR, and more likely hadronic scenarios need to be invoked. Assuming a shock on pc-scale distance from the black hole can efficiently accelerate protons, the flare could be proton induced, since the cloud provides the jet with the same amount of protons as electrons. We will elaborate on a hadronic scenario for the flare elsewhere.

The constraint on the maximum electron Lorentz factor is particularly strong from the shape of the synchrotron spectrum. Hence, the cut-off of the inverse Compton component is fixed at $\sim 20\,$GeV, which does not even take into account absorption by the EBL. If the spectrum of \cta is indeed mainly shaped by the leptonic model as described here, \cta cannot be detected at very high energy $\gamma$-rays ($E>100\,$GeV) by ground-based Cherenkov experiments. 

In summary, we have shown that the prolonged and strong activity of the FSRQ \cta could have been caused by the full or partial ablation of a gas cloud colliding with the jet. From the assumption of hydrostatic equilibrium of an isothermal gas with the gravity of the cloud, we derived the density structure of the cloud. This structure is reflected in the injection of material ablated by the jet causing the several months long outburst. Our model lightcurves are in good agreement with the observations. The model parameters suggest that the cloud was not fully ablated, and much of the material might have been lost during the collision. 

%
%
\acknowledgments
The authors are grateful for fruitful discussions with Heike Prokoph, Moritz Hackstein, and Chris Diltz, as well as for a constructive report by the anonymous referee.

The work of M.~Z. and M.~B. is supported through the South African Research Chair Initiative (SARChI) of the South African Department of Science and Technology (DST) and National Research Foundation.\footnote{Any opinion, finding and conclusion or recommendation expressed in this material is that of the authors, and the NRF does not accept any liability in this regard.}
F.~J. and S.J.~W. acknowledge support by the German Ministry for Education and Research (BMBF) through Verbundforschung Astroteilchenphysik grant 05A11VH2.
J.-P.~L. gratefully acknowledges CC-IN2P3 (\href{https://cc.in2p3.fr}{cc.in2p3.fr}) for providing a significant amount of the computing resources and services needed for this work.
A.~W. is supported by the Foundation for Polish Science (FNP).
%
%

%
\end{document}